\documentstyle[12pt,aasms4]{article}

\def\power#1{\mbox{$\times10^{#1}\ $}}

\newcommand{\msun}{M$_\odot$}
\newcommand{\msyr}{M$_\odot$ yr$^{-1}$}

\newcommand{\neo}{$^{19}$Ne}

\newcommand{\f}{$^{19}$F}

\newcommand{\otone}{$^{17}$O(p, $\gamma$)$^{18}$F(p,$\gamma$)$^{19}$Ne}

\newcommand{\al}{$^{26}$Al}

\newcommand{\mggg}{$^{26}$Mg}

\begin{document}

\title{Nucleosynthesis in Classical Novae: CO vs ONe white dwarfs}

 \author{Jordi Jos\'e}
 \affil{Departament de F\'{\i}sica i Enginyeria Nuclear (UPC), Avinguda
 V\'{\i}ctor Balaguer, s/n, E-08800 Vilanova i la Geltr\'u (Barcelona), 
 SPAIN \\ 
 and\\
 Institut d'Estudis Espacials de Catalunya (IEEC), 
 Edifici Nexus-104, \\ C/Gran Capit\`a 2-4, 08034 Barcelona, SPAIN}
 \and
 \author{Margarita Hernanz}
 \affil{Institut d'Estudis Espacials de Catalunya (IEEC), 
 CSIC Research Unit,
 Edifici Nexus-104, C/Gran Capit\`a 2-4, 08034 Barcelona, SPAIN}

\received{}
\accepted{}

\slugcomment{\underline{Submitted to}: \apj~~~~\underline{Version}:
\today}

\begin{abstract}

  Detailed nucleosynthesis in the ejecta of classical novae has been
  determined for a grid of hydrodynamic nova models. 
  The reported 14 evolutionary sequences, followed from the onset of 
  accretion up to the explosion and ejection stages, span a range of 
  CO and ONe white dwarf masses (0.8--1.35 \msun) and mixing levels 
  between the accreted envelope and the underlying white dwarf core 
  (25--75\%). 
  The synthesis of each isotope, from ${}^1$H to ${}^{40}$Ca, is 
  discussed along with its sensitivity to model parameters. 
  Special emphasis is given to isotopes such as ${}^{13}$C, ${}^{15}$N, 
  and ${}^{17}$O, which may account for a significant fraction of their 
  Galactic content. 
  Production of the radioactive isotopes ${}^7$Be, ${}^{22}$Na, and 
  ${}^{26}$Al is also analyzed, since they may provide a direct test of
  the thermonuclear runaway model through their $\gamma$-ray emission.
  The resulting elemental yields reproduce fairly well the spectroscopic
  abundance determinations of several well studied classical novae.

\end{abstract}


\section{Introduction}

Classical novae release dramatic amounts of energy.
According to the widely accepted scenario, classical novae are
close binary systems consisting of a white dwarf, and a large main 
sequence star. The companion overfills its Roche lobe and matter flows outward
through the inner Lagrangian point, leading to the formation of an 
accretion disk around the white dwarf. Some fraction of this H-rich matter 
ultimately ends up on top of the white dwarf, where it is gradually 
compressed by more material still being accreted. This compression
heats the envelope up to the point when ignition conditions to drive a
thermonuclear runaway (hereafter, TNR) are reached.

Over the last 25 years, many  hydrodynamic computations of nova 
outbursts have pointed out that an important fraction of the formerly
accreted envelope is ejected.
Since the temperatures attained in the envelope during the explosion 
are rather high,  with typical peak values of $\sim (2 - 3) \times 10^8$ K, 
the ejecta shows significant 
nuclear processing (Starrfield et al. 1997, Kovetz \& Prialnik 1997,
Jos\'e, Hernanz \& Coc 1997, and references therein).
Abundance levels of the intermediate-mass elements in the ejecta are
significantly enhanced, in general agreement with spectroscopic abundance 
determinations (Livio \& Truran 1994).
This raises the issue of the potential contribution of classical 
novae to the Galactic abundances, assuming the solar abundance levels. 
The total mass ejected by classical novae over the Galaxy's history may be
estimated by the product of the Galactic nova rate ($\sim$ 30 
events yr$^{-1}$), with the Galaxy's lifetime ($\sim 10^{10}$ yr) 
and the average 
ejected mass per nova outburst ($\sim 2 \times 10^{-5}$ \msun). 
This gives $\sim 6 \times 10^6$ \msun, which accounts
for only $\sim$ 1/3000 of the Galactic disk's gas and dust content. 
This order of magnitude estimate suggests that, despite its large occurrence 
rate in the Galaxy, novae scarcely contribute to the Milky Way's
overall metallicity, as compared with other major sources such as 
supernovae or asymptotic giant branch stars (Woosley 1986). 
However, classical novae can account for a significant fraction of the
abundance levels of individual nuclei, specially those with 
overproduction factors $f \geq 1000$
($f \equiv X_i/X_{i,\odot}$, where $X_i$ and $X_{i,\odot}$ are
the mass fractions of species $i$ in the ejected envelope and the
solar value, respectively). 
Many numerical models of nova outbursts have shown significant 
overproduction of several species, such as ${}^7$Li 
(see Starrfield, Truran \& Sparks 1978; Hernanz et al. 1996), 
${}^{13}$C, ${}^{15}$N or ${}^{17}$O 
(Starrfield et al. 1972; Sparks, Starrfield \& Truran 1976; Prialnik 1986;
 Politano et al. 1995; Kovetz \& Prialnik 1997; Starrfield et al. 1997; 
 Jos\'e \& Hernanz 1997).
In particular, Galactic ${}^{15}$N and ${}^{17}$O have been strongly supported 
as being produced during nova outbursts (see Woosley et al. 1997).
Significant production of radioactive nuclei like  ${}^{22}$Na and ${}^{26}$Al
(Weiss \& Truran 1990; Nofar, Shaviv \& Starrfield 1991;
Coc et al. 1995; Politano et al. 1995; Jos\'e, Hernanz \& Coc 1997;
Jos\'e \& Hernanz 1997; Starrfield et al. 1997), and even heavier species,
such as ${}^{31}$P, ${}^{32,33}$S or ${}^{35}$Cl, 
has also been reported (Politano et al. 1995, Jos\'e \& Hernanz 1997, 
Starrfield et al. 1997).

Despite several determinations of the nucleosynthetic yields in classical novae
are available, extended analyses
through a representative number of isotopes and nuclear reactions, 
directly linked to the hydrodynamical models, for both CO and ONe novae, 
have only been scarcely
reported. Kovetz \& Prialnik (1997) have recently published
detailed multicycle calculations of nova outbursts for CO white dwarf
masses ranging from 0.65 to 1.4 \msun. However, according 
to stellar evolution,
massive white dwarfs are expected to be made of ONe, instead of CO. 
Moreover, observations of some nova systems, such as V693 CrA 1981, 
QU Vul 1984 No. 2, V838 Her 1991, or V1974 Cyg 1992, 
reveal the presence of an underlying ONe white
dwarf. It is also worth noticing that the nuclear reactions library used 
by Kovetz \& Prialnik (1997) has been taken from Caughlan \& Fowler's (1988) 
compilation, hence, not taking into account the different updates of some
crucial rates that may influence their nucleosynthesis results.
On the other hand, Starrfield et al. (1997) have used fully updated physics
in the modeling of the neon Nova V1974 Cyg 1992. The analysis is 
limited, however, to 1.25 \msun white dwarfs, for which a fixed 50\%
degree of mixing between core and envelope is systematically adopted.

In this paper, we reinvestigate the role played by classical nova outbursts
in the synthesis of chemical species, comparing CO and ONe novae,
for a wide range of white dwarf masses (from 0.8 to 1.35 \msun) and degrees
of mixing between core and envelope (25 to 75\%), using a hydrodynamic
code with a fully updated nuclear reaction network.
Special emphasis is focused on the comparison with available observations of 
several nova systems. In Section 2 we briefly describe the method of 
computation, 
the input physics and the initial models adopted. Results from the different 
evolutionary computations and comparison with observations are given in 
Section 3. A summary of the most relevant conclusions of this paper is given 
in Section 4.

\section{Model and input physics}

Following the method described in Kutter \& Sparks (1972, 1980),
a one-dimensional, implicit, hydrodynamical code ({\it SHIVA}), in lagrangian 
formulation, has been developed to analyze the course 
of nova outbursts, from the onset of accretion up to the expansion and
ejection stages.
The code solves the standard set of differential equations for
hydrodynamical evolution: conservation of mass, momentum and energy, 
energy transport by radiation and convection, plus the definition of
the lagrangian velocity. In order to treat the long-term evolution of
the system, when the expanding nebula becomes optically thin, we have
added a term, which obeys Kirchhoff's law, to the equation of radiation 
transport in the diffusion approximation (see Larson 1972, Starrfield,
Sparks \& Truran 1974).  A time-dependent formalism for convective transport 
has been included whenever the characteristic convective
timescale becomes larger that the integration time step (Wood 1974).
Partial mixing between adjacent convective shells is treated by means of
a diffusion equation (Prialnik, Shara \& Shaviv 1979).
The equation of state includes contributions from the electron 
gas (with different degrees of degeneracy), the ion plasma and radiation;
Coulomb corrections to the electronic pressure are also taken into account. 
Models make use of Iben's (1975) opacity fits. However, the effect of 
Iglesias \& Rogers radiative opacities (1993) has also been tested. 
Calculation of evolutionary sequences including systematically the radiative
opacities from Iglesias \& Rogers (1993) is currently in progress.
The code has been linked to a reaction network, 
which follows the detailed evolution of 
100 nuclear species, ranging from $^{1}$H to $^{40}$Ca, through 
370 nuclear reactions, with updated rates and screening factors from
Graboske et al. (1973) and DeWitt, Graboske \& Cooper (1973).
The code has already been used for the specific analysis of the 
contribution of novae to the Galactic content of ${}^7$Li (Hernanz
et al. 1996) and ${}^{26}$Al (Jos\'e, Hernanz \& Coc 1997), as well as for
the gamma-ray emission from nearby novae (see Hernanz et al. 1997, 
G\'omez-Gomar et al. 1997).

Some of the input parameters with a deep influence on the nova nucleosynthesis 
are the chemical composition of the envelope and the 
mass of the underlying white dwarf. The problem 
of the chemical composition of nova envelopes is complex and far from 
being understood. Whereas Prialnik \& Kovetz (1995) strongly support the
diffusion-convection mechanism as responsible for the mixing, 
recent two-dimensional calculations by Glasner, Livne \& Truran (1997)
suggest a very efficient dredge-up of matter from the outermost
shells of the core into the solar-like accreted envelope, induced by
convection. The question
deserves further attention, probably fully three-dimensional calculations
from the onset of accretion.
 As suggested by Politano et al. (1995),
the matter transferred from the companion is assumed to be solar-like, 
and is mixed in a given fraction with the outermost shells of 
the underlying core by means of an unknown mechanism (either shear mixing, 
diffusion or a convective multidimensional process).
 This assumption is based on the enhanced CNO or ONeMg abundances 
required by theoretical nova models both to power the explosion and to 
account for the spectroscopic abundance determinations (\cite{LT94}).
In most of the calculations performed by Starrfield's group, they adopt a 
50\% degree of mixing between core and envelope. 
This may be considered as a representative mixing level, in view
of the mean metallicities observed in the ejecta of `true' ONeMg novae
(\cite{LT94}). In our models, we have considered different mixing levels
ranging from 25 to 75\%, in order to be consistent with the wide
spread of metallicities reported from observations.
 The composition of the underlying core has been taken 
from recent detailed evolutionary models, specially in the case of ONe white
dwarfs, which are the main contributors to heavy isotopes. These stars are 
made basically of $^{16}$O and $^{20}$Ne (\cite{Dom93}; \cite{Rit96}), 
since magnesium is almost absent. 
This issue plays a crucial role in the resulting nucleosynthesis, and should
be taken into account in order to compare previous estimates by different
groups. In particular, the ONeMg models by Starrfield et al. (1997)
have an initial composition of the white dwarf core based on old
nucleosynthesis calculations of C-burning from Arnett \& Truran (1969).
As stated in their paper (Starrfield et al. 1997), the use of the new
abundances by Ritossa, Garc\'\i a--Berro \& Iben (1996) may provide 
quite different results.
For the CO Models we have assumed a core composition of 
X(${}^{12}$C)=0.495, X(${}^{16}$O)=0.495, and X(${}^{22}$Ne)=0.01. 

The effect of the white dwarf mass has
been tested through a number of simulations involving both CO white dwarfs
($M_{wd}$ = 0.8, 1.0 and 1.15 \msun)
and ONe ones ($M_{wd}$ = 1.0, 1.15, 1.25 and 1.35 \msun).
The overlapping between both intervals is due to the uncertain exact 
upper (lower) limit for CO (ONe) degenerate cores.
We would like to stress that the mass accretion rate and the initial
white dwarf luminosity (or temperature) may also influence the
results. In particular, more violent outbursts are obtained when lower
mass accretion rates or lower initial luminosities are adopted, since
the higher degeneracy attained in the more massive accreted envelopes leads
to more violent outbursts. The expected effect on the
resulting nucleosynthesis is an extension of the nuclear activity 
towards heavier species as 
the mass accretion rate or the initial luminosity decrease, due to the
higher temperatures achieved in the envelope.  In this paper, we have adopted
 a mass accretion rate of $2 \times 10^{-10}$ \msyr
(despite other values have been also tested) 
and an initial luminosity of $10^{-2}$ $L_\odot$, rather
typical values, in order to limit the parameter space.

\section{Results and Discussion}

The main properties of the initial models for the 
14 evolutionary sequences presented in this paper, as well as a summary
of the most relevant characteristics of the evolution, are given in
Tables 1 and 2: the initial white dwarf mass, $M_{wd}$, 
and the adopted mixing level between core and envelope are input 
parameters; $\Delta M_{env}$ is the envelope's mass at the end
of the accretion stage; $t_{acc}$ is the duration of the accretion phase;
$t_{rise}$ is the time required for a temperature rise 
from $T_{bs} = 3 \times 10^7$ K to $10^8$ K, at the burning shell;
$\varepsilon_{nuc,max}$ and $T_{max}$ are peak nuclear energy generation
rate and maximum temperature at the burning shell, respectively;
$t_{max}$ is the time required to reach peak temperature from $T_{bs} =
10^8$ K; $\Delta M_{ejec}$, $v_{ejec}$ and $K$ represent the total
mass, the mean velocity and the mean kinetic energy of the ejected envelope.
The mean composition of the ejecta is given in Tables 3 and 4, in mass
fractions.

\subsection{Theoretical nova outbursts: from the onset of accretion to mass
ejection}

In this Section we will focus on the main characteristics
of the computed models.  As a framework for the analysis, we will describe  
the detailed evolution of Model CO5
(a 1.15 \msun CO white dwarf with a 50\% mixture with core material and
accreting mass at a rate of $2 \times 10^{-10}$ \msyr).
 
The early accretion phase is dominated by p-p chains (mainly 
${}^1$H(p,e$^+\nu_e$)${}^2$H), as well as by the CNO-cycle reaction
${}^{12}$C(p,$\gamma$)${}^{13}$N,
followed by ${}^{13}$N $\beta^+$-decay into ${}^{13}$C.
As soon as the temperature at the burning shell reaches $2.4 \times 10^7$ K, 
the nuclear timescale becomes shorter than the accretion timescale, 
and accretion becomes negligible.
The mass piled up in the envelope at the end of the accretion phase
(which lasts $\sim 10^5$ yr) is $1.8 \times 10^{-5}$ \msun. 
The rate of nuclear energy generation has risen to $\sim 10^8$ 
$\rm erg \,\, g^{-1} \, s^{-1}$.

The beginning of the TNR is accompanied by the development
of a convective zone just above the burning shell, which rapidly expands
towards the outer envelope. 
When $T_{bs}$ reaches $5 \times 10^7$ K, convection extends already
through a region of 125 km above the ignition shell.
 The release of nuclear energy
is fully dominated by the cold CNO-cycle, mainly through 
${}^{12}$C(p,$\gamma$)${}^{13}$N($\beta^+$)${}^{13}$C(p,$\gamma$)${}^{14}$N,
and no significant activity in the NeNa and MgAl-cycles is found.
A similar behavior is found at $T_{bs} = 10^8$ K, 
when convection has already extended through the whole envelope 
($\tau_{conv} \sim 0.6$ s).
The Model has spent $t_{rise} \sim 7.2 \times 10^5$ s to rise from 
$T_{bs} = 3 \times 10^7$ K to $10^8$ K. 
When temperature reaches $2 \times 10^8$ K, 
the star achieves a maximum rate of nuclear energy generation 
of $\varepsilon_{nuc,max} = 1.1 \times 10^{16}$ $\rm erg \,\, g^{-1} \, 
s^{-1}$. At this time, significant energy production comes from the hot 
CNO-cycle (initiated when ${}^{13}$N(p,$\gamma$)${}^{14}$O becomes faster 
than ${}^{13}$N($\beta^+$)${}^{13}$C), mainly through  
${}^{12}$C(p,$\gamma$)${}^{13}$N(p,$\gamma$)${}^{14}$O,
${}^{14}$N(p,$\gamma$)${}^{15}$O and ${}^{16}$O(p,$\gamma$)${}^{17}$F.
Leakage from the
MgAl-cycle through proton captures on ${}^{27}$Si and ${}^{27}$Al becomes
progressively important.
A peak temperature of $T_{max} = 2.1 \times 10^8$ K  is attained 
65 s after the ignition shell reached $10^8$ K.
As a result of the violent TNR, $1.3 \times 10^{-5}$ \msun are ejected
(72\% of the formerly accreted envelope), with a mean velocity of 
$\sim 2700$ $\rm km \, s^{-1}$ (see Table 2 for a summary of these 
results).

In order to check the effect of the white dwarf core composition, 
we have evolved Model ONe3, an ONe white dwarf 
with the same input parameters as Model CO5.
The lower amount of ${}^{12}$C present in its envelope 
reduces the role played by the CNO-cycle and
less nuclear energy is released at the same temperature. 
Therefore, Model ONe3 accretes a more massive envelope before the TNR
begins ($3.2 \times 10^{-5}$ \msun).
Since the ignition density (and, hence, the degeneracy) is also higher, 
a higher peak temperature is attained ($T_{max} = 2.2 \times 10^8$ K).
The net effect is a partial extension of the nuclear activity towards
higher $Z$ nuclei, both because of the different peak temperature and the
different chemical composition of the envelope. In particular,  Model ONe3
shows the dominant role played by some reactions of the MgAl-cycle at peak
temperature, which are absent in Model CO5.
A second feature, which turns out to be crucial, is
the different timescales of the TNR: Model ONe3 requires $t_{rise}
\sim 1.3 \times 10^7$ s to increase the temperature at the burning shell from
$T_{bs} = 3 \times 10^7$ K to $10^8$ K, plus $t_{max} \sim 540$ s to reach
peak value (see Table 1 for a summary of the results). These larger times 
deeply influence the final abundances in the ejecta (see Section 3.2).

In order to mimic the uncertain process of mixing between the solar-like
accreted material and the outermost layers of the underlying CO or ONe white
dwarf, we have adopted different degrees of mixing ranging from 25 to 75\%. 
Computations with $1.15$ \msun ONe white dwarfs (i.e., Models ONe2, ONe3 and 
ONe4) show that a more massive envelope is accreted when a higher degree
of mixing is adopted, leading to a more violent outburst. For 
instance, Model ONe4 (with 75\% mixing) attains a peak temperature
of $T_{max} = 2.5 \times 10^8$ K and ejects matter
with a mean kinetic energy of $K = 1.9 \times 10^{45}$ erg, as 
compared with Model ONe2 (with only 25\% mixing), for which 
$T_{max} = 2.2 \times 10^8$ K and  $K = 1.1 \times 10^{45}$ erg 
(see Table 1).
A similar trend is found for Models ONe6 and ONe7, involving 1.35 \msun
ONe white dwarfs with 50 and 75\% mixing, respectively.
We have also performed several computations
involving 1.15 \msun CO white dwarfs (i.e., Models CO4, CO5 and CO6, with
25, 50 and 75\% mixing, respectively), as well as 0.8 \msun CO white dwarfs
(Models CO1 and CO2, with 25 and 50\%, respectively). Contrary to the ONe
Models, the most massive envelopes are accreted on top of white dwarfs
with 25\% mixing, with a minimum mass around 50\% mixing. However, the 
strength of the explosion, as indicated by a higher peak temperature
and a higher mean kinetic energy, increases with the mixing level
(see Table 2).

As shown in Tables 1 and 2, massive white dwarfs develop a TNR
after a shorter accretion phase (and hence, accreting less mass)
as compared with lighter white dwarfs, because of the higher surface gravity.
Also the evolution towards peak temperature takes place with a shorter
timescale. 
The most relevant outcome is the increase of the peak
temperature attained during the TNR as the mass of the white dwarf increases.
We stress that this is specially noticeable for Model ONe6 (with $M_{wd} = 
1.35$ \msun), which attains a maximum temperature of $3.2 \times 10^8$ K.

Two different prescriptions for the radiative opacities 
have been considered in order to estimate
their potential effect on the progress of the outburst and on the
resulting nucleosynthesis: Model CO5 has been evolved using Iben's (1975) 
fits to the opacity tables of Cox \& Stewart (1970a, b), whereas 
Iglesias \& Rogers opacities (1993) have been adopted in Model CO7.  
As shown by Jos\'e (1996), the use of Iglesias \& Rogers (1993) opacities
reduces the mass of the accreted envelope, leading to a softer explosion.
The reason is that Iglesias \& Rogers opacities are larger 
than Iben's ones. Therefore, a significant temperature increase in the
envelope of Model CO7 ensues, reducing the time required to achieve the 
critical conditions for a TNR (see Table 2). A similar trend has been 
recently pointed out
by Starrfield et al. (1997).  Nucleosynthesis results from Model
CO7 do not reveal large differences with those from Model CO5 
(see Table 4).

\subsection{Nucleosynthesis}

In this Section, we will examine the yields obtained in our numerical 
nova models. Tables 3 and 4 list the mean chemical composition of the ejecta,
in mass fractions, few days after the explosion,
resulting from our evolutionary sequences of ONe and CO novae, 
respectively. Overproduction factors, relative to
solar abundances, for Models CO5, ONe3 and ONe6, are displayed in Figures
1 to 3.

\subsubsection{Synthesis of ${}^7$Be--${}^7$Li}

The synthesis of ${}^7$Li in classical novae has been recently analyzed
in detail by Hernanz et al. (1996), who have confirmed that the 
{\it beryllium transport} mechanism can efficiently lead to
large amounts of ${}^7$Li. In that paper, we showed that
lithium production is favored when CO novae, instead of ONe 
ones, are adopted. The faster evolution of CO novae allow
photodisintegration of ${}^8$B through ${}^8$B($\gamma$,p)${}^7$Be to
prevent ${}^7$Be destruction (synthesized in the first part of the TNR 
by means of ${}^3$He($\alpha,\gamma$)${}^7$Be).

Ejected masses of ${}^7$Li in the CO Models are almost an order of 
magnitude larger than in the ONe ones, with a maximum production for a 
50\% mixing.  Because of the higher degeneracy attained in massive white 
dwarfs, which results in stronger outbursts with shorter evolutionary 
timescales, production is enhanced when the initial mass of the underlying 
white dwarf is increased.
Despite large overproduction factors are obtained for most of the CO Models 
(up to $f \sim 900$, see Fig. 1), classical novae can only account for 
$\sim$ 10\% of the Galactic ${}^7$Li content, assuming the solar system level. 
This result is similar to the estimates given by Starrfield, Truran \& Sparks 
(1978), in the framework of hydrodynamic models of CO nova outbursts, but 
assuming an initial envelope already in place (hence, neglecting the accretion 
phase and the building-up of the envelope). In their most recent hydrodynamic
nova models (Starrfield et al. 1997), they obtain lower overproduction
factors of ${}^7$Li than the ones reported from our evolutionary sequences. 
This is probably due to the different choice of initial 
 abundances, and, to some extent, to other differences in the input physics 
(such as reaction rates and equation of state) or even in the treatment of 
convection. Since CO novae dominate ${}^7$Li synthesis, other calculations 
involving CO novae are needed in order to compare them with our results. 
In the recent analysis of CO novae by Kovetz \& Prialnik (1997), the
limited nuclear network, ranging from ${}^{12}$C to ${}^{31}$P, does not
enable any study of light elements.

${}^7$Be has another 
potential interest as a gamma-ray signature of a nearby nova outburst 
(Clayton 1981, Harris, Leising \& Share 1991), since 
its decay to ${}^7$Li, with the emission of a gamma-ray photon of 478 keV, 
may be detected for CO novae within 0.5 kpc by the future INTEGRAL mission 
(Hernanz et al. 1997, G\'omez-Gomar et al. 1997). 

\subsubsection{Synthesis of the CNO-group nuclei}

The dominant nuclear reaction at the beginning of the TNR in a nova
outburst is typically ${}^{12}$C(p,$\gamma$)${}^{13}$N, which is followed by a 
combination of $\beta^+$-decays, or (p,$\gamma$) and (p,$\alpha$) reactions, as 
a function of the local temperature.
As pointed out by Starrfield et al. (1972), some of the most overabundant
species at peak temperature, except hydrogen and helium, are the short
lived, $\beta^+$-unstable nuclei
${}^{13}$N, ${}^{14}$O, ${}^{15}$O and ${}^{17}$F, which decay releasing
enough energy to account for the ejection of a fraction of the envelope.
Therefore, their daughter nuclei ${}^{13}$C, ${}^{14, 15}$N and ${}^{17}$O 
are among the main products in the ejecta of classical novae.
Models presented in this paper show large overproduction factors
of ${}^{13}$C, ${}^{15}$N, and ${}^{17}$O, with respect to solar abundances
(see Figs. 1--3).

The synthesis of ${}^{13}$C is initiated by ${}^{12}$C(p,$\gamma$)${}^{13}$N.
Its evolution follows a competition between destruction through 
${}^{13}$C(p,$\gamma$)${}^{14}$N near the burning shell, and production by 
means of ${}^{13}$N($\beta^+$)${}^{13}$C, in the outer 
envelope, where a fraction of ${}^{13}$N has been carried out by convection.
Our computations show that the synthesis of ${}^{13}$C is favored in CO novae,
due to the higher initial content of ${}^{12}$C. Hence, 
its final amount increases when higher degrees of mixing with the 
underlying core are adopted. Overproduction factors for the different CO 
Models computed lay in the range $\sim 900 - 5200$, with a
maximum abundance of X(${}^{13}$C) $\sim$ 0.2, by mass. Overproduction factors 
ranging from 400 to 900 are obtained in the ONe Models.

The final amount of ${}^{15}$N is related to the fraction of its parent 
nucleus ${}^{15}$O (coming from ${}^{12}$C(p,$\gamma$)${}^{13}$N($\beta^+$)
${}^{13}$C(p,$\gamma$)${}^{14}$N(p,$\gamma$)${}^{15}$O) that is 
transported to the outer layers by convection before decaying.
Its evolution follows a competition between destruction by proton captures 
(basically the closing CNO-cycle reaction ${}^{15}$N(p,$\alpha$)${}^{12}$C)
and creation through ${}^{15}$O($\beta^+$)${}^{15}$N.
${}^{15}$N is overproduced for both CO (f $\sim 200 - 9200$) and
ONe Models (f $\sim 1800 - 32000$). The higher abundances of
${}^{15}$N found in the ONe Models result from the higher peak temperatures
achieved, which allow proton captures to proceed onto ${}^{14}$N. 
Only some CO Models show overproduction factors of ${}^{14}$N larger than 100
($M_{wd} \sim 0.8 - 1.15$ \msun, with 50\% - 75\% mixing). 

The synthesis of ${}^{17}$O is dominated by  
${}^{16}$O(p,$\gamma$)${}^{17}$F($\beta^+$)${}^{17}$O, whereas destruction
is due to both ${}^{17}$O(p,$\alpha$)${}^{14}$N and
${}^{17}$O(p,$\gamma$)${}^{18}$F (only at high temperatures).
The last reaction is in turn responsible for the synthesis of ${}^{18}$O from 
 ${}^{18}$F($\beta^+$)${}^{18}$O.  Our calculations show a significant 
overproduction of ${}^{17}$O: $f \sim 900 - 5500$ in the CO Models, 
and $f \sim 2900 - 13000$ in the ONe ones.
The increase of the final abundances with the white dwarf mass and also 
when ONe cores are adopted is a direct consequence of the higher peak 
temperatures achieved, which allow proton captures to proceed onto ${}^{16}$O 
and initiate the chain.

Accurate estimates of the contribution of classical novae to the Galactic
abundances of these CNO-group nuclei (or of any other species) require a
model of chemical evolution of the Galaxy, which properly takes them into
account. Present models of chemical evolution include novae in a rather
rough way, without taking into account the wide range of variation of the
yields with nova properties (see, for instance, Woosley et al. 1997, in 
which the 1.35 \msun ONeMg Model from Politano et al. 1995 is adopted as 
representative).
This can be partially due to the fact that these yields were not available 
until very recently (i.e., Kovetz \& Prialnik 1997, for CO novae, and the
present work for CO and ONe novae). A detailed analysis is out of the scope
of this paper, but a crude estimate of the Galactic production of some
elements by novae can be obtained from our 
evolutionary sequences. We have 
derived upper limits to this production which may be useful in order to
elucidate which elements deserve a careful analysis and which ones can be
discarded as being produced by novae. Upper limits are obtained from the
following expression:
$$\rm M^{i} (M_\odot) = \tau_G (yr) \times M_{ej}^i (M_\odot) \times  
                        \nu_{nova} (yr^{-1})$$
where $\rm \tau_G$ is the age of the Galaxy, $\rm \nu_{nova}$ is the Galactic 
nova rate and $\rm M_{ej}^i$ corresponds to the ejected mass of species $i$
in the most favorable model. We have adopted $\rm \tau_G \sim 10^{10}$ yr,
and $\rm \nu_{nova} \sim 30 \, yr^{-1}$ (Shafter 1997).

Our estimates show that nova outbursts may account
for the 100\% of the Galactic abundances of ${}^{13}$C, ${}^{15}$N and 
${}^{17}$O, assuming solar system levels for the mean composition
of the Milky Way. However, since the maximum production of ${}^{15}$N 
is attained for massive ONe novae (which are less abundant than low-mass 
CO novae, 
according to stellar evolution), this upper limit is probably too high.
As expected, novae scarcely contribute to the abundances
of ${}^{14}$N and ${}^{18}$O (with upper limits around 10 and 25\% of the 
Galactic abundances, respectively). 
As a summary, classical novae are likely sites for the synthesis
of most (or all) of the Galactic ${}^{13}$C and ${}^{17}$O, and may 
contribute significantly to the amount of ${}^{15}$N, though an extra
source seems to be required.
 These results are in good agreement with the main conclusions addressed
 by Kovetz \& Prialnik (1997) in their analysis of the composition in
 the ejecta of CO novae. 

 Another important feature obtained in our calculations is that the ratios 
 O/N and C/N decrease when the mass of the white dwarf increases.
 For instance, in our ONe Models, the ratio O/N ranges from 3.3 (for a 
 1.15 \msun white dwarf) to 0.2 (1.35 \msun), whereas C/N lays between
 1.1 (1.15 \msun) and 0.2 (for a 1.35 \msun white dwarf). Hence, high
 concentrations of N in the ejecta of a nova system may reveal the
 presence of a massive white dwarf. Our results also indicate that a decrease
 in the degree of mixing between core and envelope, from 75 to 25\%,
 translates into lower ratios O/N. No clear influence on C/N is found in
 the ONe Models.
 The observed trend agrees with results from previous works (Politano et al.
 1995, for ONeMg novae; Kovetz \& Prialnik 1997, for CO ones).
 
\subsubsection{Synthesis of ${}^{19}$F}

The nucleosynthetic origin of \f, by far the least abundant of all the 
stable $12 \leq A \leq 35$ nuclei, is still a matter of debate. 
A handful of astrophysical scenarios has been suggested, including 
explosive hydrogen-burning sites like classical novae 
(Woosley 1986; \cite{Wie86}; \cite{Tru86}),
 thermal pulses in AGB stars
(Forestini et al. 1992; Jorissen, Smith \& Lambert 1992;
 Mowlavi, Jorissen \& Arnould 1996),
Wolf-Rayet stars (Meynet \& Arnould 1993), 
proton-ingestion into He-rich layers (Jorissen \& Arnould 1989),
 and neutrino process during type II supernovae (Woosley \& Haxton
1988; Timmes, Woosley \& Weaver 1995; Woosley \& Weaver 1995).

The mechanism responsible for the synthesis of ${}^{19}$F in classical novae 
is the production of the short-lived, $\beta^+$-unstable nucleus ${}^{19}$Ne
through \otone,  
which is partially transported by convection towards the outer, cooler layers 
of the envelope where it decays into \f, with a mean lifetime of
 $\tau$(\neo) $\sim$ 25 s.
Our calculations show that ${}^{19}$F production 
is more important in ONe novae than in CO ones. It
increases as the white dwarf mass and the degree of mixing 
increase (see Table 3), with overproduction factors 
$\sim 100$ obtained for Model ONe7 (a 1.35 
\msun white dwarf with 75\% mixing). 
These values are not far
from the ones found during thermal pulses in asymptotic giant branch stars
(see \cite{For92}; \cite{Jor92}). 
Nevertheless, since ${}^{19}$F is only significantly synthesized in models
involving 1.35 \msun white dwarfs (see Fig. 3), which are very scarce in 
nature, we conclude that classical novae
can be ruled out as dominant sources of the Galactic \f.
The importance of ${}^{19}$F lays on the fact that it 
provides a potential observational clue of the presence of a massive
white dwarf.  

\subsubsection{Synthesis of the NeNa and MgAl-group nuclei}

Two isotopes the NeNa and MgAl-groups have a particular astrophysical interest:
${}^{26}$Al and ${}^{22}$Na. These nuclei are synthesized in significant amounts
in ONe rather than in CO novae. 

${}^{26}$Al is an unstable nucleus, with a lifetime of $\tau = 1.04 \times 
10^6$ years, that decays from ground state to the first excited level of \mggg,
which in turn de-excites to its ground state by emitting a 
gamma-ray photon of 1809 keV.
This characteristic gamma-ray signature of \al, first detected by 
the HEAO-3 satellite (\cite{Mah82}, \cite{Mah84}), has been 
confirmed by other space missions like SMM      
(\cite{Sha85}) and several balloon-borne experiments.
Recent measurements made with the COMPTEL 
instrument on-board CGRO have provided 
a map of the 1809 keV emission in the Galaxy (Diehl et al. 1995; 
Prantzos \& Diehl 1996). 
They derive a total ${}^{26}$Al mass between 1 and 3 \msun which, 
according to the
observed distribution, is mainly attributed to young progenitors, such as 
massive AGB stars, type II supernovae and Wolf-Rayet stars.
However, a potential contribution from novae or low-mass AGB stars has not been
excluded. 
The synthesis of ${}^{26}$Al in classical novae has been analyzed
in a recent paper by Jos\'e, Hernanz \& Coc (1997), who have shown
that only some combinations of peak temperatures around T$_{\rm{peak}} 
\sim$ 1--2\power{8} K and rapid evolution from maximum favor ${}^{26}$Al
generation in ONe novae. 
A crude estimate of the total production of ${}^{26}$Al 
brings a maximum contribution of classical nova outbursts 
 in the range 0.1 to 0.4 \msun. Despite the contribution of the yields from 
Model ONe1
(i.e., 1.00 \msun white dwarf with 50\% mixing) would increase this range
drastically, we point out that, according to the results from
the theory of stellar evolution, 1.00 \msun white dwarfs are expected to
be made of CO, instead of ONe. Therefore, Model ONe1 should be considered
as a test model, rather than a representative model for ONe novae. 

The synthesis of ${}^{22}$Na proceeds through 
${}^{20}$Ne(p,$\gamma$)${}^{21}$Na
(starting from either initial ${}^{20}$Ne or from ${}^{23}$Na, that can be
transformed in the former by ${}^{23}$Na(p,$\alpha$)${}^{20}$Ne),
which can either decay into ${}^{21}$Ne (at low temperature), followed by
${}^{21}$Ne(p,$\gamma$)${}^{22}$Na, or capture another proton leading to
${}^{21}$Na(p,$\gamma$)${}^{22}$Mg($\beta^+$)${}^{22}$Na 
(at high temperature). Since ${}^{20}$Ne requires temperatures around
$T \sim 4 \times 10^8$ K to burn, only massive white dwarfs, which attain
higher peak temperatures close to that value allow an efficient synthesis of 
${}^{22}$Na.  Our models show that its final amount increases when the
white dwarf mass or the degree of mixing with the core increase, in
agreement with the anticorrelation between ${}^{22}$Na and ${}^{26}$Al 
production reported by Politano et al. (1995). 

${}^{22}$Na has a lifetime of $\tau\sim 3.75$ years. In its decay to a 
short-lived excited state of ${}^{22}$Ne, it emits a gamma-ray photon of
1275 keV (Clayton \& Hoyle, 1974). A few experimental verifications of 
the gamma-ray emission at 
1275 keV from nearby novae have been attempted in the last decades,
using balloon-borne experiments and several detectors on-board HEAO-3,
SMM and CGRO, from which constraints on the maximum amount of ${}^{22}$Na 
ejected into
the interstellar medium have been derived (Leising et al. 1988, Harris et 
al. 1996). The most recent estimates are 
based on several measurements with the COMPTEL experiment on-board CGRO of 
five recent Ne-novae (Iyudin et al. 1995),
which translate into an upper limit of the 
ejected ${}^{22}$Na mass by any nova in the Galactic disk of the order of
$\rm 3.7 \times 10^{-8}$ \msun.  The low ejected masses of ${}^{22}$Na 
obtained in our ONe Models agree fairly
well with this upper limit.  Moreover, our estimated gamma-ray flux from a 
nearby nova (i.e., within 0.5 kpc) is too low to be detected by OSSE or 
COMPTEL, but may represent a good target for the future INTEGRAL mission 
(Hernanz et al. 1997, G\'omez-Gomar et al. 1997). 

${}^{22}$Na has also been invoked to account for the abnormally 
high amount of ${}^{22}$Ne in the Ne-E meteorites (Clayton 1975; Eberhardt
et al. 1979; Lewis et al. 1979) in which the extremely high content
of Ne (the possibility of pure ${}^{22}$Ne is not excluded)
is attributed to the $\beta^+$-decay of ${}^{22}$Na.
However, since the shells ejected during ONe nova outbursts are contaminated 
 with large amounts of ${}^{20}$Ne (and some ${}^{21}$Ne)
 the ratio (${}^{20}$Ne + ${}^{21}$Ne)/(${}^{22}$Ne + ${}^{22}$Na) 
is much larger than 1; thus, it seems
 unlikely to explain the isotopic ratios found in the Ne-E meteorites,
 despite some hypothetical mechanisms such as elemental fractionation
 have been invoked to reduce the ratio down to the observed values.
The problem of the isotopic ratio 
 (${}^{20}$Ne + ${}^{21}$Ne)/(${}^{22}$Ne + ${}^{22}$Na) has already been
pointed out by Hillebrandt \& Thielemann (1982) and by Wiescher et al. 
(1986), who also obtained ratios higher than 1 in 
  their nucleosynthesis calculations with parametrized nova models.

 Another isotope of the MgAl cycle, ${}^{27}$Al, is also significantly
synthesized during ONe nova outbursts. It is produced basically by means of
${}^{26}$Mg(p,$\gamma$)${}^{27}$Al.  Near peak temperature, creation through
${}^{27}$Si($\beta^+$)${}^{27}$Al dominates the destruction channel 
${}^{27}$Al(p,$\gamma$)${}^{28}$Si. However, above $2 \times 10^8$ K,
proton captures on ${}^{27}$Si dominate its $\beta^+$-decay and, hence, only a 
rapid evolution from peak temperature avoids ${}^{27}$Al destruction. 
The maximum production is attained in Model ONe1 (i.e., 1.0 \msun 
white dwarf), which achieves the lowest peak temperature, 
$T_{max} = 1.98 \times 10^8$ K, and also in Model ONe4 (i.e., 1.15
\msun white dwarf with 75\% mixing), which shows a fast decline 
from peak temperature.
Even in the most favorable cases, contribution of novae to the Galactic 
${}^{27}$Al is limited to less than 15\%. 

\subsubsection{Synthesis of heavier species: from Si to Ca}

The nuclear activity in this range of isotopes is basically restricted to 
massive ONe white dwarfs, which 
achieve high enough temperatures to enable proton captures onto heavy nuclei.
The most overproduced species are the odd nuclei ${}^{31}$P, 
${}^{33}$S and ${}^{35}$Cl (see Fig. 3).
 Similar results were also obtained by Starrfield et al. (1997) 
 (See their Model 8, with 1.35 \msun). 
 The significant synthesis of ${}^{35}$Cl obtained in Models ONe6 and ONe7
 confirms the estimates given by Woosley (1986) with parametrized 
 calculations. Significant detection of Cl in the ejecta of a nova outburst
 would therefore reveal the presence of a very massive ONe white dwarf.
 It is unlikely that classical novae may play any role in the Galactic 
abundances of these elements: even the most favorable case accounts for 
less than 20\% of the Galactic ${}^{33}$S and $\sim$ 10\% of 
 ${}^{35}$Cl and ${}^{31}$P.

 It is worth noticing that a recent spectrum of the dwarf nova VW Hydri, 
 taken with the Hubble Space Telescope  (Sion et al. 1997), has revealed
 the presence of P with an abundance 900 times solar,
 not far from the values obtained in Model ONe6. This observation has 
 raised the 
 question if a thermonuclear runaway has occurred on that dwarf nova.
 Quite surprisingly, the estimated mass of the white dwarf in the nova
 system VW Hydri is around 0.86 \msun, too low to allow proton captures
 to proceed up to ${}^{31}$P. 

\subsection{The ejecta of classical novae: a comparison with observations}

The comparison between observed and theoretical abundances is not 
straightforward for two reasons. On one hand, completely different yields 
have been derived for some systems, as for instance V693 CrA 1981 and 
QU Vul 1984.  This fact points out the intrinsic difficulties in the 
accurate determination of the chemical abundances from observations.
Up to which extent the reported abundances are precisely known is somehow 
uncertain.
On the other hand, one should bear in mind that numerical calculations lack
from an accurate description of some input physics: the 
mechanism responsible for metal enrichment of the accreted 
envelope, a self-consistent formulation of convection, and realistic 
prescriptions for the adopted nuclear reaction rates, to quote a few.

In Table 5 we show the comparison between spectroscopic abundance 
determinations (in mass fractions) and those obtained from our models, 
for five novae which are particularly well fited.
Model ONe3 (i.e., a $1.15$ \msun white dwarf, with 50\% 
mixing), shows similar yields to those derived for the nova 
system V693 CrA 1981, according to the recent reanalysis by 
Vanlandingham et al. (1997),
despite the intrinsic differences in the mean metallicity.
The presence of an underlying ONe white dwarf can be inferred from 
the high amount of Ne in the ejecta, as well as  from 
the moderately high concentration of nuclei in the range Na-Fe,
$\sim 0.06$ by mass. Mass fractions of H, C, N, Ne and Na-Fe agree quite well 
with the observed values (see Table 5). 
The largest difference is the final amount of O (with a mass fraction of 
$\sim 0.07$ derived from observations, as compared with $\sim 0.15$ 
obtained in Model ONe3). Our value is a direct consequence of mixing 
with the outer layers of an ONe white dwarf extremely enriched in ${}^{16}$O,
according to the adopted chemical composition of Ritossa, 
Garc\'\i a--Berro \& Iben (1996). 
The second main difference is the helium 
content: V693 CrA 1981 has an abnormally high concentration of this element
that, as suggested by Starrfield et al. (1997), may result from residual 
H-burning in the remaining shells of a previous outburst. 
Other spectroscopic abundance determinations of V693 CrA 1981 
(Andre\"a et al. 1994, Williams et al. 1985) provide completely different 
results. In fact, the abundance distribution derived by Andre\"a et al. 
(1994), with nearly twice the metallicity obtained by 
Vanlandingham et al. (1997), is better fited by Model ONe4
(i.e., a $1.15$ \msun white dwarf with 75\% mixing). 
Furthermore, Model ONe5 roughly reproduces the distribution derived by
Williams et al. (1985), with an excellent agreement in H, N, O and Ne. 

Another nova system, V1370 Aql 1982, shows low amounts of H and He and a 
high concentration of Ne, in an envelope with a mean metallicity of 
$\sim 0.86$ (Andre\"a et al. 1994). 
This suggests a high mixing level with the Ne-rich shells of a massive white 
dwarf.
Model ONe7, a $1.35$ \msun white dwarf with 75\% mixing, fits appreciably
well the derived abundances (see Table 5). 
It is worth noticing that  Snijders et al. (1987) inferred a high concentration
of sulfur in the ejecta of V1370 Aql 1982, 
a determination that is still a matter of debate. We point out that similar
amounts of sulfur are obtained in Models ONe6 and ONe7.

The abundance distribution derived for another `neon' nova, QU Vul 1984 
(Austin et al. 1996), is roughly similar to the mean composition obtained
in Model ONe1 (i.e., a 1.0 \msun white dwarf with 50\% mixing). 
This recent determination widely differs from previous estimates 
by Andre\"a et al. (1994) and Saizar et al. (1992).

Model CO4 (a $1.15$ \msun white dwarf with 25\% mixing) also fits fairly well 
the composition derived for the classical nova PW Vul 1984
(Andre\"a et al. 1994). The lack of heavy nuclei in the ejecta, specially
Ne, suggests the presence of a low mass CO white dwarf, a conclusion also
supported by the moderately high amount of C in the ejecta. 
The total amount of nuclei in the range Na-Fe is slightly lower than the 
value derived from observations. This is a consequence of the adopted 
initial
chemical composition, since the moderate temperatures achieved in low mass 
white dwarfs do not allow
significant nuclear flows towards this range of isotopes. 
Other abundance distributions for nova PW Vul 1984 have been obtained by
different groups: Saizar et al. (1991) derived a much lower metallicity 
($\sim 0.067$, only twice solar), whereas the recent reanalysis 
by Schwarz et al. (1997), using another photoionization code, 
has provided a distribution that lays between the 
estimates by Saizar et al. (1991) and Andre\"a et al. (1994). In 
particular, the new determination of the C content is reduced 
with respect to the value obtained by Andre\"a et al. (1994).

 Moreover, Model CO4 also fits the chemical distribution derived for 
another nova system, V1688 Cyg 1978 (Andre\"a et al. 1994), where the 
most relevant discrepancy is the absence of nuclei in the range Na-Fe.

\section{Conclusions}

We have computed 14 hydrodynamic models of nova outbursts, 
from the onset of accretion up to the ejection 
stage, for a range of CO and ONe white dwarf masses (0.8 to 1.35 
\msun), and degrees of mixing between the accreted envelope and the 
outermost shells of the underlying white dwarf core (25 to 75\%).
The main characteristics of the explosions as well as a detailed 
nucleosynthesis for all of them are provided. These yields can be
important for future detailed studies of the chemical evolution of the 
Galaxy which intend to include novae in an accurate way.

The role played by the different model parameters has been analyzed. 
Concerning the influence of the composition of the underlying white 
dwarf, we obtain that ONe novae accrete more massive envelopes than CO ones 
(if all the remaining input parameters are the same), due to the lower 
$^{12}$C content in the envelope when the TNR develops. As a result of 
the higher degeneracy, a more violent outburst is found. Therefore, ONe 
models show a partial extension of the nuclear activity towards higher $Z$ 
nuclei, due to the higher peak temperature achieved in their envelopes.

More violent outburts also occur as the mass of the white dwarf increases, 
because of the larger surface gravity. The most relevant outcome is the 
synthesis of heavy nuclei, such as P, S or Cl, as a result of the higher 
peak temperature attained in the envelope. However, since massive white 
dwarfs are very scarce in nature, their potential contribution to the 
Galactic abundances is rather small, despite the large overproduction 
factors of some particular isotopes.

In order to reproduce the wide spread of metallicities reported from 
observations, a range of mixing levels between the core and the envelope 
has been adopted. The general trend is an increase of peak temperature and 
mean ejection velocities as the degree of mixing increases. Higher 
degrees of mixing favor the synthesis of higher $Z$ nuclei in ONe models, 
whereas a clear pattern is not found for the CO novae.

We have shown that classical novae are likely sites for the synthesis of
most (or all) of the Galactic ${}^{13}$C and ${}^{17}$O (with maximum 
overproduction factors $\sim$ 5200 and $\sim$ 13000, respectively), 
and may also contribute significantly to the abundance of ${}^{15}$N
($f_{\rm max} \sim$ 32000), though an extra source seems to be required.
CO Models produce significant amounts of ${}^7$Be ($f_{\rm max} \sim$ 900), 
large enough to
be detected from nearby novae through its $\gamma$-ray emission, providing a 
potential observational clue of the presence of a CO white dwarf. 
Contribution of classical novae to the Galactic $^7$Li, coming from $^7$Be 
decay, is limited to $\sim$ 10\%.

The ejecta from ONe Models show an important synthesis of ${}^{22}$Na.
Its $\gamma$-ray emission might be detected by future space missions 
according to the values obtained in our calculations which are in good 
agreement with the upper limits derived from COMPTEL observations. 
Concerning $^{26,27}$Al, ONe novae can account for less than 10-15\% of the 
Galactic content.

Our nova models show that massive ONe white dwarfs are characterized by low 
O/N and C/N ratios in the ejecta. Also, the presence of a massive ONe white 
dwarf could be inferred from a significant detection of ${}^{19}$F, 
${}^{35}$Cl, and even ${}^{31}$P and ${}^{33}$S in ejected nova shells.

The elemental yields obtained in our grid of nova models fit fairly
well the spectroscopic abundance determinations of the novae V693 CrA 1981, 
V1370 Aql 1982, QU Vul 1984, PW Vul 1984 and V1688 Cyg 1978.  

\acknowledgments
We are grateful to the referee, Francis X. Timmes, for many valuable 
suggestions that have greatly improved the presentation of this paper. 
This research has been partially supported by the DGICYT (PB94-0827-C02-02), 
by the CICYT (ESP95-0091), and by the CIRIT (GRQ94-8001).

\newpage

\begin{deluxetable}{cccccccc}
\tablewidth{0 pt}
\tablecaption{Initial parameters and main characteristics of 
ONe nova Models}
\tablehead{
\colhead{}&
\colhead{ONe1}&
\colhead{ONe2}&
\colhead{ONe3}&
\colhead{ONe4}&
\colhead{ONe5}&
\colhead{ONe6}&
\colhead{ONe7}
}
\startdata
$M_{wd}$ (\msun)              & 1.00& 1.15& 1.15& 1.15& 1.25& 1.35& 1.35\nl
\% Mixing                     & 50  & 25  & 50  & 75  & 50  & 50  & 75\nl
$\Delta M_{env}$ ($10^{-5}$ \msun)
                              & 6.4 & 3.2 & 3.2 & 3.5 & 2.2 & 0.54& 0.58\nl
$t_{acc}$ ($10^5$ yr)         & 3.3 & 1.9 & 1.9 & 2.1 & 1.3 & 0.31& 0.33\nl
$t_{rise}$ ($10^6$ s)         & 20  & 46  & 13  & 11  & 6.8 & 2.5 & 2.1\nl
$\varepsilon_{nuc,max}$ ($10^{16}$ $\rm erg \, g^{-1} \, s^{-1}$)
                              & 0.29& 0.36& 0.76 & 2.4& 2.1 & 19  & 14 \nl
$T_{max}$ ($10^8$ K)          & 1.98& 2.21& 2.19& 2.48& 2.44& 3.24& 3.32\nl
$t_{max}$ (s)                 & 768 & 828 & 540 & 305 & 380 & 150 & 108 \nl
$\Delta M_{ejec}$ ($10^{-5}$ \msun) 
                              & 4.7 & 2.3 & 1.9 & 2.6 & 1.4 & 0.44& 0.34\nl
$v_{ejec}$ ($\rm km \, s^{-1}$)
                              & 1600& 2100& 2400 & 2500 & 3100& 4100& 6000 \nl
$K$ ($10^{45}$ erg)           & 1.3 & 1.1 & 1.2 & 1.9 &  1.4 & 0.9 & 1.3\nl
\enddata
\end{deluxetable}

\newpage

\begin{deluxetable}{cccccccc}
\tablewidth{0 pt}
\tablecaption{Initial parameters and main characteristics of 
CO nova Models}
\tablehead{
\colhead{}&
\colhead{CO1}&
\colhead{CO2}&
\colhead{CO3}&
\colhead{CO4}&
\colhead{CO5}&
\colhead{CO6}&
\colhead{CO7 \tablenotemark{a}}
}
\startdata
$M_{wd}$ (\msun)       & 0.8 & 0.8 & 1.0 & 1.15& 1.15& 1.15& 1.15\nl
\% Mixing              & 25  & 50  & 50  & 25  & 50  & 75  & 50  \nl
$\Delta M_{env}$ ($10^{-5}$ \msun) 
                       & 9.7 & 8.8 & 3.9 & 2.1 & 1.8 & 1.8 & 0.9 \nl
$t_{acc}$ ($10^5$ yr)  & 3.1 & 2.6 & 1.7 & 1.2 & 1.0 & 1.0 & 0.4 \nl
$t_{rise}$ ($10^6$ s)  & 2.8 & 1.8 & 1.1 & 0.43& 0.72& 0.48& 1.7 \nl
$\varepsilon_{nuc,max}$ ($10^{16}$ $\rm erg \, g^{-1} \, s^{-1}$)
                       &0.05 & 0.1 & 0.3 & 0.5 & 1.1 & 1.9 & 0.2 \nl
$T_{max}$ ($10^{8}$ K) & 1.45& 1.51& 1.70& 2.03& 2.05& 2.08& 1.73\nl
$t_{max}$ (s)          & 454 & 199 & 152 & 147 &  65 & 51  & 200 \nl
$\Delta M_{ejec}$ ($10^{-5}$ \msun)
                       & 7.0 & 6.4 & 2.3 & 1.5 & 1.3 & 1.3 & 0.63\nl
$v_{ejec}$ ($\rm km \, s^{-1}$)    
                       & 800 & 1200& 1900& 2200& 2700& 2900& 2700\nl
$K$ ($10^{45}$ erg)    & 0.6 & 1.1 & 0.9 & 0.8 & 1.0 & 1.3 & 0.45\nl
\enddata
\tablenotetext{a}{Model with Iglesias \& Rogers opacities (1993)}
\end{deluxetable}

\newpage

\begin{deluxetable}{cccccccc}
\tablecolumns{8}
\tablewidth{0 pt}
\footnotesize
\tablecaption{Yields from ONe nova Models (Mass Fractions)}
\tablehead{
\colhead {} & \multicolumn{7}{c}{Models} \\
\cline{2-8} \\
\colhead{Nucleus}&
\colhead{ONe1}&
\colhead{ONe2}&
\colhead{ONe3}&
\colhead{ONe4}&
\colhead{ONe5}&
\colhead{ONe6}&
\colhead{ONe7}
}
\startdata
 $^1$H     &3.2E-1 &4.7E-1  &3.0E-1 & 1.2E-1 & 2.8E-1 & 2.4E-1 & 7.3E-2\nl
 $^3$He    &7.1E-8 &2.1E-9  &4.3E-8 & 1.7E-7 & 2.8E-8 & 2.9E-8 & 9.7E-8\nl
 $^4$He    &1.8E-1 &2.8E-1  &2.0E-1 & 1.3E-1 & 2.2E-1 & 2.4E-1 & 1.7E-1\nl
 $^7$Be    &2.3E-7 &4.6E-8  &6.0E-7 & 1.2E-6 & 6.9E-7 & 1.3E-6 & 2.4E-6\nl
 $^{11}$B  &8.8E-13&4.0E-13 & 4.4E-12& 1.7E-11& 1.2E-11& 2.5E-10& 1.9E-9\nl
 $^{12}$C  &1.3E-2 &1.8E-2  &2.3E-2 & 2.2E-2 & 2.8E-2 & 2.1E-2 & 2.6E-2\nl
 $^{13}$C  &1.7E-2 &2.3E-2  &2.8E-2 & 2.7E-2 & 3.2E-2 & 1.5E-2 & 2.5E-2\nl
 $^{14}$N  &2.6E-2 &3.0E-2  &2.2E-2 & 2.7E-2 & 3.2E-2 & 4.6E-2 & 3.5E-2\nl
 $^{15}$N  &7.7E-3 &1.7E-2  &2.3E-2 & 2.4E-2 & 4.2E-2 & 1.2E-1 & 1.4E-1\nl
 $^{16}$O  &1.7E-1 &2.4E-2  &1.2E-1 & 2.3E-1 & 7.1E-2 & 2.2E-2 & 9.1E-2\nl
 $^{17}$O  &1.8E-2 &1.1E-2  &2.8E-2 & 4.1E-2 & 3.9E-2 & 1.7E-2 & 5.1E-2\nl
 $^{18}$O  &8.2E-3 &2.4E-3  &6.0E-3 & 7.3E-3 & 4.2E-3 & 9.8E-4 & 1.8E-3\nl
 $^{19}$F  &8.5E-6 &4.7E-6  &8.9E-6 & 1.2E-5 & 1.3E-5 & 2.2E-5 & 4.0E-5\nl
 $^{20}$Ne &1.8E-1 &9.0E-2  &1.8E-1 & 2.6E-1 & 1.8E-1 & 1.5E-1 & 2.4E-1\nl
 $^{21}$Ne &1.9E-5 &1.3E-5  &3.0E-5 & 4.0E-5 & 3.5E-5 & 5.1E-5 & 8.4E-5\nl
 $^{22}$Ne &2.0E-3 &5.9E-4  &1.7E-3 & 2.5E-3 & 1.0E-3 & 1.5E-4 & 4.2E-4\nl
 $^{22}$Na &4.8E-5 &3.1E-5  &5.3E-5 & 1.5E-4 & 9.6E-5 & 6.0E-4 & 6.5E-4\nl
 $^{23}$Na &1.2E-3 &3.6E-4  &7.5E-4 & 3.6E-3 & 1.4E-3 & 6.6E-3 & 7.9E-3\nl
 $^{24}$Mg &2.5E-4 &1.6E-5  &1.0E-4 & 1.5E-3 & 2.0E-4 & 3.6E-4 & 1.2E-3\nl
 $^{25}$Mg &1.0E-2 &7.8E-4  &2.9E-3 & 7.4E-3 & 2.4E-3 & 4.2E-3 & 6.6E-3\nl
 $^{26}$Mg &9.4E-4 &7.8E-5  &3.4E-4 & 1.0E-3 & 2.8E-4 & 5.9E-4 & 1.0E-3\nl
 $^{26}$Al &2.7E-3 &1.8E-4  &9.3E-4 & 2.0E-3 & 5.4E-4 & 7.2E-4 & 1.5E-3\nl
 $^{27}$Al &1.2E-2 &7.6E-4  &4.5E-3 & 9.2E-3 & 2.0E-3 & 1.8E-3 & 4.5E-3\nl
 $^{28}$Si &3.4E-2 &3.0E-2  &5.4E-2 & 7.3E-2 & 5.6E-2 & 3.5E-2 & 5.8E-2\nl
 $^{29}$Si &8.7E-5 &3.1E-4  &4.2E-4 & 7.8E-4 & 8.8E-4 & 1.7E-3 & 2.7E-3\nl
 $^{30}$Si &4.3E-5 &1.4E-3  &6.9E-4 & 1.7E-3 & 4.8E-3 & 1.1E-2 & 1.7E-2\nl
 $^{31}$P  &4.5E-6 &2.6E-4  &5.9E-5 & 1.9E-4 & 1.3E-3 & 7.6E-3 & 1.2E-2\nl
 $^{32}$S  &2.0E-4 &3.6E-4  &2.0E-4 & 1.2E-4 & 8.3E-4 & 2.3E-2 & 1.9E-2\nl
 $^{33}$S  &4.7E-6 &4.3E-5  &1.2E-5 & 7.0E-6 & 7.7E-5 & 9.1E-3 & 4.4E-3\nl
 $^{34}$S  &9.2E-6 &1.8E-5  &9.2E-6 & 4.7E-6 & 1.9E-5 & 6.4E-3 & 1.8E-3\nl
 $^{35}$Cl &1.5E-6 &6.2E-6  &2.2E-6 & 1.2E-6 & 6.1E-6 & 7.0E-3 & 8.7E-4\nl
 $^{36}$S  &4.6E-8 &5.4E-8  &4.2E-8 & 2.1E-8 & 3.2E-8 & 5.4E-9 & 5.7E-9\nl
 $^{36}$Ar &3.9E-5 &5.8E-5  &3.9E-5 & 1.9E-5 & 3.8E-5 & 3.9E-3 & 1.9E-4\nl
 $^{37}$Cl &4.8E-7 &1.4E-6  &6.2E-7 & 3.4E-7 & 1.2E-6 & 2.8E-4 & 7.2E-6\nl
 $^{38}$Ar &7.7E-6 &1.1E-5  &7.6E-6 & 3.8E-6 & 7.4E-6 & 5.1E-5 & 3.7E-6\nl
 $^{39}$K  &1.8E-6 &2.9E-6  &1.8E-6 & 9.1E-7 & 2.0E-6 & 6.5E-6 & 1.8E-6\nl
\enddata
\end{deluxetable}

\newpage
\begin{deluxetable}{cccccccc}
\tablecolumns{8}
\tablewidth{0 pt}
\footnotesize
\tablecaption{Yields from CO nova Models (Mass Fractions)}
\tablehead{
\colhead {} & \multicolumn{7}{c}{Models} \\
\cline{2-8} \\
\colhead{Nucleus}&
\colhead{CO1}&
\colhead{CO2}&
\colhead{CO3}&
\colhead{CO4}&
\colhead{CO5}&
\colhead{CO6}&
\colhead{CO7 \tablenotemark{a}}
}
\startdata
 $^1$H     & 5.1E-1 & 3.3E-1 & 3.2E-1 & 4.7E-1 & 3.0E-1 & 1.2E-1 & 3.0E-1 \nl
 $^3$He    & 7.0E-6 & 9.2E-6 & 6.1E-6 & 1.5E-6 & 4.1E-6 & 2.8E-6 & 3.7E-6 \nl
 $^4$He    & 2.1E-1 & 1.4E-1 & 1.5E-1 & 2.5E-1 & 1.6E-1 & 9.0E-2 & 1.6E-1 \nl
 $^7$Be    & 4.4E-7 & 9.6E-7 & 3.1E-6 & 6.0E-6 & 8.1E-6 & 4.0E-6 & 3.1E-6 \nl
 $^{11}$B  &1.1E-13 &2.2E-14 &1.7E-12 & 2.6E-11& 2.2E-11& 7.4E-12& 1.9E-12\nl
 $^{12}$C  & 1.4E-2 & 5.3E-2 & 3.6E-2 & 2.9E-2 & 4.8E-2 & 6.8E-2 & 3.2E-2 \nl
 $^{13}$C  & 3.4E-2 & 1.1E-1 & 1.3E-1 & 4.4E-2 & 9.6E-2 & 1.9E-1 & 1.0E-1 \nl
 $^{14}$N  & 9.5E-2 & 1.1E-1 & 1.1E-1 & 7.1E-2 & 1.1E-1 & 1.4E-1 & 1.4E-1 \nl
 $^{15}$N  & 9.9E-4 & 9.3E-4 & 6.2E-3 & 2.3E-2 & 4.0E-2 & 2.9E-2 & 5.5E-3 \nl
 $^{16}$O  & 1.3E-1 & 2.5E-1 & 2.4E-1 & 8.6E-2 & 2.1E-1 & 3.4E-1 & 2.3E-1 \nl
 $^{17}$O  & 3.3E-3 & 4.4E-3 & 8.0E-3 & 1.2E-2 & 2.1E-2 & 1.9E-2 & 8.6E-3 \nl
 $^{18}$O  & 8.4E-4 & 5.6E-4 & 2.2E-3 & 4.4E-3 & 3.8E-3 & 3.7E-3 & 3.9E-3 \nl
 $^{19}$F  & 8.5E-7 & 4.4E-7 & 9.9E-7 & 5.0E-6 & 3.4E-6 & 1.8E-6 & 1.7E-6 \nl
 $^{20}$Ne & 1.2E-3 & 8.2E-4 & 8.5E-4 & 1.4E-3 & 9.7E-4 & 5.2E-4 & 8.7E-4 \nl
 $^{21}$Ne & 2.9E-8 & 4.0E-8 & 5.6E-8 & 1.9E-7 & 1.7E-7 & 7.2E-8 & 3.4E-8 \nl
 $^{22}$Ne & 2.6E-3 & 5.0E-3 & 5.0E-3 & 2.2E-3 & 4.8E-3 & 7.3E-3 & 5.0E-3 \nl
 $^{22}$Na & 3.4E-7 & 3.0E-7 & 1.6E-7 & 3.8E-7 & 2.9E-7 & 1.1E-7 & 8.5E-8 \nl
 $^{23}$Na & 3.6E-5 & 3.6E-5 & 3.4E-5 & 1.6E-5 & 2.0E-5 & 2.4E-5 & 3.4E-5 \nl
 $^{24}$Mg & 5.7E-5 & 6.3E-5 & 1.6E-5 & 4.4E-6 & 1.8E-5 & 1.0E-5 & 2.8E-6 \nl
 $^{25}$Mg & 3.8E-4 & 2.4E-4 & 2.8E-4 & 1.1E-4 & 1.6E-4 & 1.3E-4 & 2.8E-4 \nl
 $^{26}$Mg & 5.5E-5 & 3.7E-5 & 3.0E-5 & 1.1E-5 & 1.5E-5 & 9.4E-6 & 2.6E-5 \nl
 $^{26}$Al & 8.1E-6 & 3.4E-6 & 1.6E-5 & 3.1E-5 & 4.7E-5 & 3.3E-5 & 2.4E-5 \nl
 $^{27}$Al & 4.8E-5 & 2.6E-5 & 4.3E-5 & 1.3E-4 & 1.3E-4 & 5.9E-5 & 5.4E-5 \nl
 $^{28}$Si & 4.9E-4 & 3.3E-4 & 3.3E-4 & 9.4E-4 & 4.5E-4 & 1.9E-4 & 3.4E-4 \nl
 $^{29}$Si & 2.6E-5 & 1.7E-5 & 1.7E-5 & 1.6E-5 & 1.3E-5 & 7.3E-6 & 1.7E-5 \nl
 $^{30}$Si & 1.8E-5 & 1.2E-5 & 1.2E-5 & 3.2E-5 & 1.7E-5 & 7.5E-6 & 1.2E-5 \nl
 $^{31}$P  & 6.1E-6 & 4.1E-6 & 4.1E-6 & 6.2E-6 & 4.0E-6 & 2.0E-6 & 4.0E-6 \nl
 $^{32}$S  & 3.0E-4 & 2.0E-4 & 2.0E-4 & 2.9E-4 & 2.0E-4 & 9.8E-5 & 2.0E-4 \nl
 $^{33}$S  & 2.5E-6 & 1.7E-6 & 1.8E-6 & 8.6E-6 & 3.3E-6 & 1.3E-6 & 1.9E-6 \nl
 $^{34}$S  & 1.4E-5 & 9.3E-6 & 9.3E-6 & 1.4E-5 & 9.2E-6 & 4.6E-6 & 9.3E-6 \nl
 $^{35}$Cl & 1.9E-6 & 1.3E-6 & 1.3E-6 & 2.4E-6 & 1.4E-6 & 6.7E-7 & 1.3E-6 \nl
 $^{36}$S  & 7.0E-8 & 4.7E-8 & 4.7E-8 & 6.8E-8 & 4.6E-8 & 2.3E-8 & 4.7E-8 \nl
 $^{36}$Ar & 5.8E-5 & 3.9E-5 & 3.9E-5 & 5.8E-5 & 3.9E-5 & 1.9E-5 & 3.9E-5 \nl
 $^{37}$Cl & 6.4E-7 & 4.3E-7 & 4.3E-7 & 7.4E-7 & 4.6E-7 & 2.2E-7 & 4.3E-7 \nl
 $^{38}$Ar & 1.2E-5 & 7.7E-6 & 7.7E-6 & 1.2E-5 & 7.7E-6 & 3.8E-6 & 7.7E-6 \nl
 $^{39}$K  & 2.6E-6 & 1.7E-6 & 1.7E-6 & 2.6E-6 & 1.7E-6 & 8.7E-7 & 1.7E-6 \nl
\enddata
\tablenotetext{a}{Model with Iglesias \& Rogers opacities (1993)}
\end{deluxetable}

\newpage

\newpage
\begin{deluxetable}{ccccccccc}
\footnotesize
\tablecaption{Models vs observations of some classical nova systems} 
\tablewidth{0 pt}
\tablehead{
\colhead{}&
\colhead{H}&
\colhead{He}&
\colhead{C}&
\colhead{N}&
\colhead{O}&
\colhead{Ne}&
\colhead{Na-Fe}&
\colhead{Z}
}
\startdata
V693 CrA 1981            &    &    &      &     &     &    &       &     \nl
\cline{1-1}
Vanlandingham et al. 1997&0.25&0.43&0.025 &0.055&0.068&0.17& 0.058 & 0.32\nl
Model ONe3               &0.30&0.20&0.051 &0.045&0.15 &0.18& 0.065 & 0.50\nl 
\cline{1-1}
Andre\"a et al. 1994     &0.16&0.18&0.0078&0.14 &0.21 &0.26& 0.030 & 0.66\nl
Model ONe4               &0.12&0.13&0.049 &0.051&0.28 &0.26& 0.10  & 0.75\nl 
\cline{1-1}
Williams et al. 1985     &0.29&0.32&0.0046&0.080&0.12 &0.17& 0.016 & 0.39\nl
Model ONe5               &0.28&0.22&0.060 &0.074&0.11 &0.18& 0.071 & 0.50\nl 
\cline{1-9}
V1370 Aql 1982           &    &    &      &     &     &    &       &     \nl
\cline{1-1}
Andre\"a et al. 1994    &0.044&0.10&0.050 &0.19 &0.037&0.56& 0.017 & 0.86\nl
Model ONe7              &0.073&0.17&0.051 &0.18 & 0.14&0.24& 0.14  & 0.76\nl 
\cline{1-1}
Snijders et al. 1987    &0.053&0.088&0.035&0.14 &0.051&0.52& 0.11  & 0.86\nl
Model ONe7              &0.073&0.17&0.051 &0.18 & 0.14&0.24& 0.14  & 0.76\nl 
\cline{1-9}
QU Vul 1984              &    &    &      &     &     &    &       &     \nl
\cline{1-1}
Austin et al. 1996       &0.36&0.19&      &0.071& 0.19&0.18& 0.0014& 0.44\nl
Model ONe1               &0.32&0.18& 0.030&0.034& 0.20&0.18& 0.062 & 0.50\nl
\cline{1-1}
Saizar et al. 1992       &0.30&0.60&0.0013&0.018&0.039&0.040&0.0049& 0.10\nl
Model ONe2               &0.47&0.28& 0.041&0.047&0.037&0.090&0.0035& 0.25\nl
\cline{1-9}
PW Vul 1984              &    &    &      &     &     &    &       &     \nl
\cline{1-1}
Andre\"a et al. 1994     &0.47&0.23&0.073 & 0.14&0.083&0.0040&0.0048&0.30\nl
Model CO4                &0.47&0.25&0.073 &0.094& 0.10&0.0036&0.0017&0.28\nl
\cline{1-9}
V1688 Cyg 1978           &    &    &      &     &     &    &       &     \nl
\cline{1-1}
Andre\"a et al. 1994     &0.45&0.22&0.070 &0.14 & 0.12&      &      &0.33\nl
Model CO4                &0.47&0.25&0.073 &0.094& 0.10&0.0036&0.0017&0.28\nl
\cline{1-1}
Stickland et al. 1981   &0.45&0.23&0.047 &0.14 & 0.13&0.0068&      &0.32\nl
Model CO1                &0.51&0.21&0.048 &0.096& 0.13&0.0038&0.0015&0.28\nl
\enddata
\end{deluxetable}

\clearpage
\figcaption[jjosefig1.eps]{
   Overproduction factors, relative to solar abundances, versus mass
   number for Model
   CO5 (a 1.15 \msun CO white dwarf with 50\% mixing).}  

\figcaption[jjosefig2.eps]{
   Same as Fig. 1, for Model
   ONe3 (a 1.15 \msun ONe white dwarf with 50\% mixing).}  

\figcaption[jjosefig3.eps]{
   Same as Fig. 1, for a 1.35 \msun ONe white dwarf with 50\% mixing
   (Model ONe6).}  

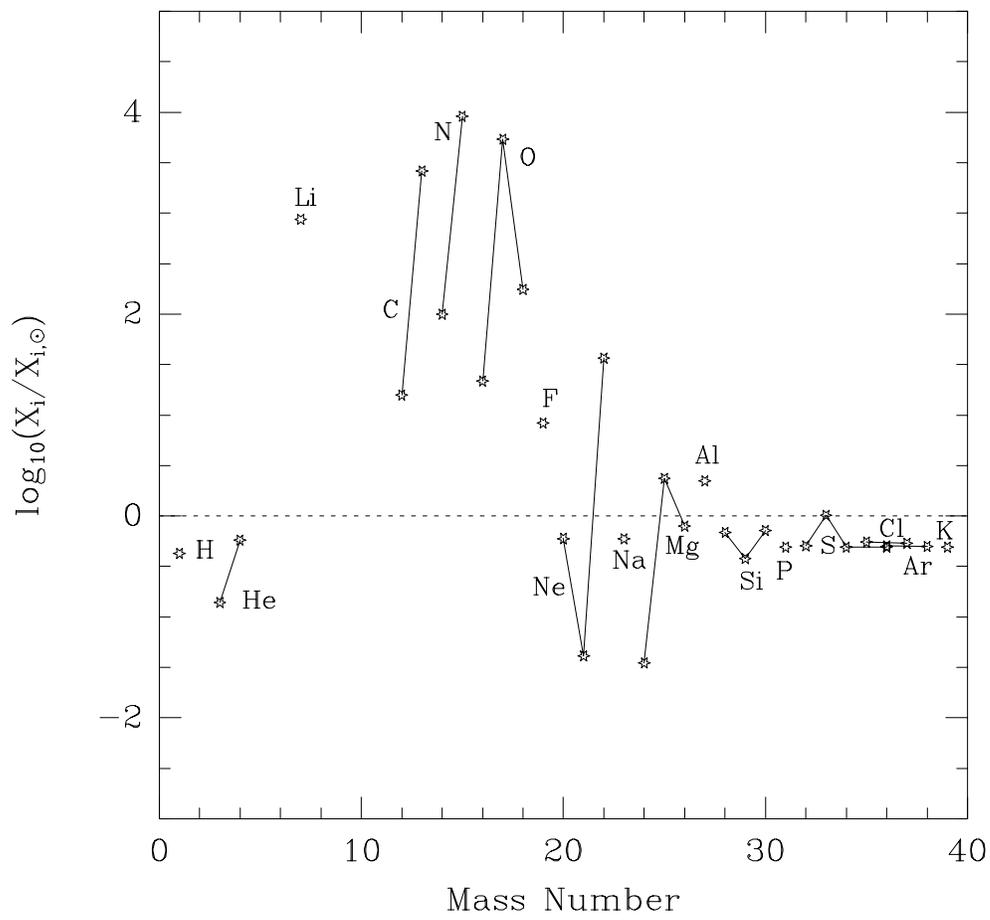
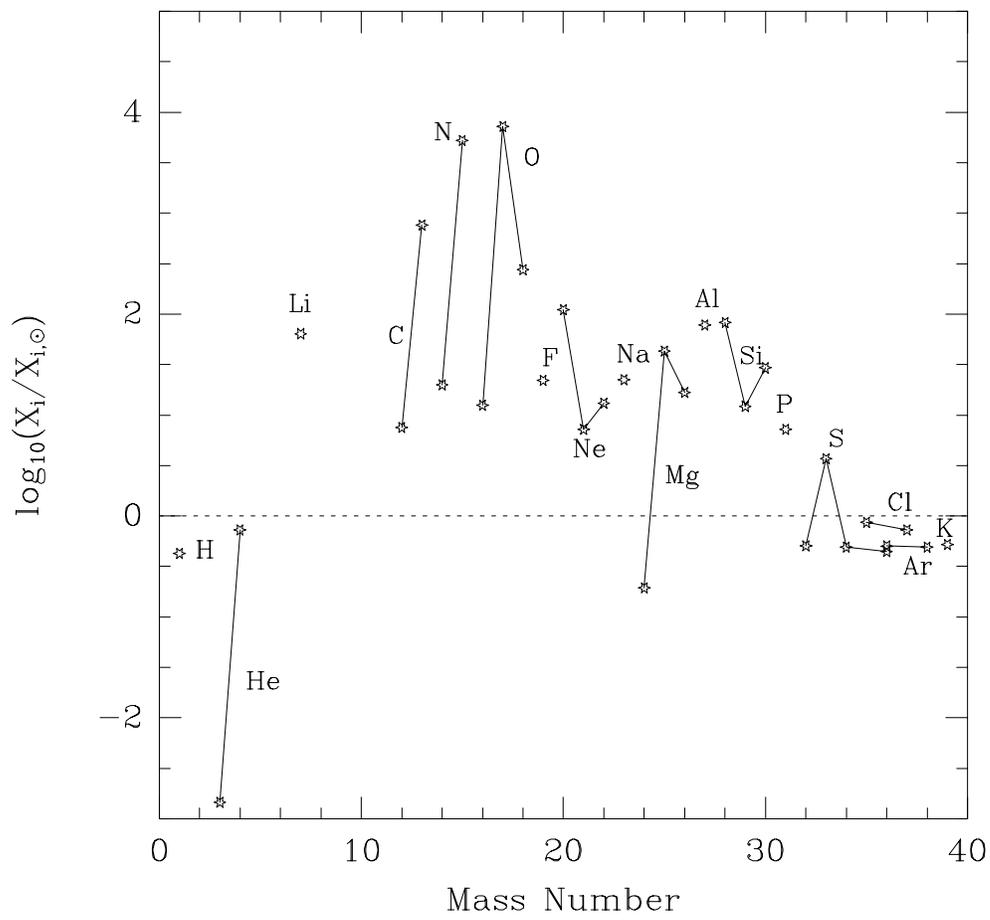
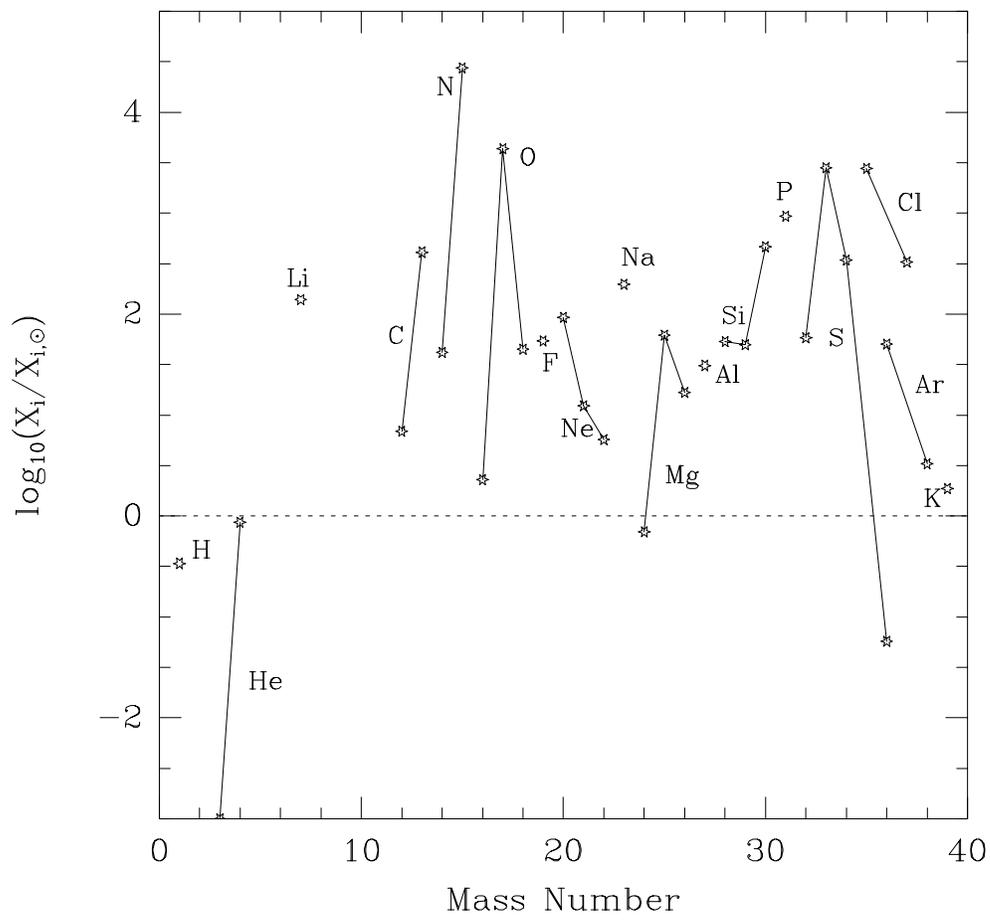
\end{document}